\documentclass[preprint,12pt,authoryear]{elsarticle}

\usepackage{amssymb}
\usepackage{amsmath}
\usepackage{lipsum}
\usepackage{graphicx}
\usepackage{xcolor}
\usepackage{dcolumn}
\usepackage{bm}
\usepackage{slashed}
\usepackage{hyperref}

\journal{Nuclear Physics B}

\begin{document}

\begin{frontmatter}



\title{A Likelihood Ratio Framework for Highly Motivated Subdominant Signals}  


\author[first]{S. Ansarifard}
\ead{ansarifard@ipm.ir}
\affiliation[first]{organization={School of physics, Institute for Research in Fundamental Sciences (IPM)}, \\
            postcode={P.O.Box 19395-5531},
            city={Tehran},
            country={Iran}}
\begin{abstract}
In particle physics and cosmology, distinguishing subtle new physics signals from established backgrounds is a fundamental and persistent challenge for phenomenologists. This paper discuss a simple and robust statistical framework to evaluate the compatibility of highly motivated (HM) theoretical models with the residuals of experimental results, focusing on scenarios where the data appear consistent with background predictions. A likelihood ratio test is developed that compares null and alternative hypotheses, emphasizing cases where new physics introduces small deviations from the background. The practicality of the framework is highlighted, and in addition to its limitations, strategies to simplify complex background modeling are discussed.
\end{abstract}




\end{frontmatter}




\section{Introduction}
\label{introduction}
The outcome of many experiments often result in a failure to reject the null hypothesis, meaning their results are compatible with the predictions of standard physics, collectively referred to the background. However, motivated by various theoretical and experimental considerations, phenomenologists search for signs of new physics within these results. In many cases, the signal of new physics is expected to manifest as a subtle fluctuation over the well-understood background. In this regard, two key issues arising: a) the false identification of features and b) the concealment of critical information that statistical analysis can reveal. The first issue is a topic frequently discussed in the literature, where random fluctuations or systematic effects can be exaggerated, leading to the perception of false signals. This is one of the primary reasons why the 5$\sigma$ level of significance is required for discoveries in particle physics \cite{Junk:2020azi}. By contrast, we explore cases where weak or hidden signals, are uncovered through statistical tools. This is particularly important, as potential future discoveries may lie among phenomena with $\sim 2\sigma - 3\sigma$ significance. Consequently, it is crucial to develop robust and straightforward procedures to study such scenarios, which are often considered as hint or tension. 

Analyzing experimental results presents numerous challenges and requires careful consideration, demanding both skill and experience. For comprehensive yet brief reviews on statistics in particle physics, see \cite{Cousins:2018tiz}, while \cite{Lyons:2017zlp} provides an insightful example of delicate but crucial details that require attention. However, researchers often need rapid yet reliable estimations to determine whether further investigation is warranted. This is more important when there is a theoretical motivation that supports our belief in the alternatives. As a result, for one seeking to evaluate the potential of new physics without delving deeply into complex analysis, the likelihood ratio test within the classical Frequentist hypothesis test serves as an excellent starting point. This is a well-established method that has been extensively studied \citep{Cowan:2010js, Blennow:2013oma}. Nevertheless, robust statistical inference demands meticulous attention to detail. Key challenges include but not limited to accurately estimating the tail distributions of p-values to enhance precision \cite{Fowlie:2021gmr}, properly accounting for effects such as the \textit{look-elsewhere} \citep{Davies:1987zz, Ranucci:2005ep, Gross:2010qma} and incorporating the systematic \cite{vanDyk:2023tqz}. These considerations are essential for drawing ultimate conclusions from experimental data.

Hypothesis testing provide measure to compares the compatibility of two models with experimental measurements. These models are defined by the null hypothesis ($H_0$) and alternative ($H_1$) hypothesis, each with its own set of free parameters (for standard definitions, see Chap. 40 in \cite{ParticleDataGroup:2024cfk}). The null hypothesis $H_0$ represents a well-established model with strong empirical belief, while the alternative $H_1$ proposes a theoretically motivated but less constrained scenario. Crucially, we distinguish between ``belief'', grounded in replicated experimental results and ``motivation", driven by anomalies in observations, theoretical extensions, or exploratory curiosity for detecting new phenomena in the future. To reflect this distinction, we label $H_0$ as Strongly Believed (SB) and $H_1$ as Highly Motivated (HM). 

$ H_0 $ and  $ H_1 $ are typically modeled with $ p_0 $ and $ p_1 $ free parameters, respectively. In many theories, $ H_0 $ is nested within $ H_1 $, meaning $ H_0 $ can be derived from $ H_1 $ by fixing $ \tilde{p}_1 $ parameters. Here, $ \tilde{p}_1 = p_1 - p_0 $ represents the number of additional parameters introduced by new physics. If $ H_1 $ includes parameters not shared with $ H_0 $ or beyond the $ \tilde{p}_1 $ constrained parameters, the look-elsewhere effect must be accounted for. In this study, we exclude such parameters. According to Wilks' theorem \cite{Wilks:1938dza}, for a sufficiently large sample size, the likelihood ratio test statistic $ \mathrm{t} $ under $ H_0 $ follows a chi-square distribution with $ \tilde{p}_1 $ degrees of freedom. It is important to emphasize that the alternative hypothesis $H_1$ carries a risk of overfitting unless it is genuinely HM that is discussed in details in following section. In such situations, hypothesis testing remains a valuable tool for uncovering potential new signals. In this work, we aim to clarify the conditions under which the outcome of a hypothesis test can be regarded as reliable. The next section details the derivation and interpretation of this test.

\section{Likelihood Ratio Test}
\subsection{Statistical Framework}
Let $\mathbf{x} = \{x_i\}_{i=1}^{n}$ denote a set of $n$ measurements, each associated with a label $t_i$ and an experimental uncertainty $\sigma_i$.  
In typical high-energy physics applications, $x_i$ represents a count-based observable, though it may also correspond to other reconstructed summary statistics. The label $t_i$ is an independent parameter, such as energy or time bin. The uncertainties $\sigma_i$ quantify the total estimated experimental error, derived either from numerical simulation or, where an analytic likelihood exists, from likelihood-based fitting procedures. In general, the full uncertainty model is described by an $n \times n$ covariance matrix. However, in many practical analyses, correlated components can be incorporated into the summary statistics via suitable rearrangements of the measurement space or through the introduction of nuisance parameters into the theoretical prediction $\bm{\mu}$.

The goal of the analysis is to test the compatibility of a theoretical prediction $\bm{\mu}$—which may include nuisance parameters—with the observed data $\mathbf{x}$.  
To distinguish between the underlying population and a particular observed sample, we denote population-level quantities with a tilde: for example, $\mathbf{\tilde{x}}$ refers to the generic random variables for a set of measurement, whereas $\mathbf{x}$ represents its observed realization. Similarly, $\bm{\mu}$ denotes the nominal prediction, which may differ from the true underlying value $\tilde{\bm{\mu}}$. We assume that each measurement $x_i$ is drawn from a Gaussian distribution with mean $\tilde{\mu}_i$ and standard deviation $\tilde{\sigma}_i$, where $\tilde{\mu}_i$ and $\tilde{\sigma}_i$ are the unknown true values.

\medskip
To quantify the level of agreement between the theoretical prediction $\bm{\mu}$ and the observed data $\mathbf{x}$, we use the $\chi^2$ test statistic
\begin{equation}\label{eq:chi2test}
    \chi^2 = \sum_{i=1}^{n} \frac{(x_i - \mu_i)^2}{\sigma_i^2}.
\end{equation}
The statistic $\chi^2$ is itself a random variable. Its distribution can be understood by rewriting Eq.~\eqref{eq:chi2test} in the form
\begin{equation}\label{eq:chi2rewrite}
    \chi^2 = \sum_{i=1}^{n} w_i^2 \big( z_i - \lambda_i \big)^2,
\end{equation}
where the weights $w_i = \tilde{\sigma}_i / \sigma_i$ reflect the ratio of the true uncertainty to its estimated value,
\[
    z_i = \frac{x_i - \tilde{\mu}_i}{\tilde{\sigma}_i},
\]
follows a standard normal distribution $\mathcal{N}(0,1)$, and
\[
    \lambda_i = \frac{\mu_i - \tilde{\mu}_i}{\tilde{\sigma}_i}
\]
quantifies the normalized deviation of the nominal prediction $\mu_i$ from the true value $\tilde{\mu}_i$.  
Collectively, $\bm{\lambda} = (\lambda_1,\dots,\lambda_n)$ measures the ability of the data to discriminate between the model under test $\bm{\mu}$ and the hypothetical true model $\tilde{\bm{\mu}}$.

The estimated uncertainties $\sigma_i$ should be close to or larger than the true $\tilde{\sigma}_i$, implying $w_i \lesssim 1$.  
Values of $w_i$ larger than unity indicate an underestimate of the uncertainty, which artificially enhances the contribution of the corresponding measurement.  

The exact distribution of $\chi^2$ in Eq.~\eqref{eq:chi2rewrite} does not, in general, have a closed form.  
However, if we approximate $w_i \approx 1$, then $\chi^2$ reduces to a sum of squares of independent normal variables with means $\lambda_i$ and unit variance.  
In this case, $\chi^2$ follows a non-central chi-square distribution $\chi^2_n(s)$ with $n$ degrees of freedom and non-centrality parameter 
\begin{equation}\label{eq:s}
    s = \sum_{i=1}^{n} \lambda_i^2,
\end{equation}
often referred to as the \emph{signal strength}.  
When the nominal prediction coincides with the truth ($\bm{\mu} = \tilde{\bm{\mu}}$), one obtains $s = 0$ and the distribution reduces to the standard (central) chi-square distribution with $n$ degrees of freedom.  

A special case of practical interest is $\tilde{\bm{\mu}} = \mathbf{0}$, which corresponds to testing whether the prediction $\bm{\mu}$ can be distinguished from pure statistical fluctuations in the absence of a true signal.

\medskip
In practice, the model prediction $\bm{\mu}$ is typically obtained by fitting to the data, which yields a best-fit point where the $\chi^2$ statistic is minimized. With  $\bm{\mu}$ is given as a true model, the value at this minimum, denoted $\chi^2_{\text{min}}$, is expected to follow a chi-square distribution with $n-p$ degrees of freedom, where $n$ is the number of data points and $p$ the number of fitted parameters.

A widely used diagnostic for assessing the goodness of fit is the \emph{standardized deviation} of the minimum chi-square. Considering the chi-square distribution with mean $n-p$ and variance $2(n-p)$, for $n-p \gtrsim 5$, one may define
\begin{equation}
Z = \frac{\chi^2_{\min} - (n-p)}{\sqrt{2(n-p)}} .
\end{equation}
This quantity measures the deviation of the observed $\chi^2_{\min}$ from its expected value in units of standard deviation. Values in the range $-1 < Z < 1$ indicate a satisfactory fit. Values $Z < -2$ may indicate overfitting. Conversely, $Z > 3$ shows a poor fit and suggests that the model should be rejected. In the intermediate regime,
$2 < Z < 3$, the model is not strictly excluded, but the fit quality is generally considered unsatisfactory. This is interesting region which may point to unaccounted systematics, statistical fluctuations, or the presence of new physics. This statistical framework will serve as the basis for defining the Strongly Believed (SB) Null Hypothesis in the following section.

\subsection{SB Hypothesis}
Under the null hypothesis $H_0$, the observed data are described by a background model
$\bm{\mu} = \{b_i(\boldsymbol{\eta})\}_{i=1}^{n}$, where $\boldsymbol{\eta} = \{\eta_\alpha \}_{\alpha=1}^{p_0}$ denotes the free parameters of $H_0$.
We define a null hypothesis as SB if it satisfies three conditions: I) The background model must provide an acceptable fit to the data. Under the null hypothesis, the chi-square statistic is
\begin{equation}\label{eq:chimin0}
\chi^2(\boldsymbol{\eta}) = \sum_{i=1}^n \frac{\big(x_i - b_i(\boldsymbol{\eta})\big)^2}{\sigma_i^2},
\end{equation}
minimizing this expression yields the best-fit parameter value $\hat{\boldsymbol{\eta}}$. We require that the minimum chi-square lies within an approximate acceptance region while remaining sufficiently far from a trivial perfect fit. For a SB null hypothesis, we impose $Z \lesssim 3$. II) The background model at best fit value must be statistically well-defined and distinguishable from pure noise. This requirement can be expressed as a condition on $b_i(\hat{\boldsymbol{\eta}})$ in such a way that
\begin{equation}\label{eq:signalst}
\sum_i \frac{b_i^2}{\sigma_i^2} \gtrsim 6 \sqrt{2(n-p)},
\end{equation}
This condition ensure the probability of getting $Z \lesssim 3$ while the data is drawn from pure noise is less than $\sim \%2$. This can be shown by considering the fact that mean and variance of non central chi square distribution follows $n - p + s$ and $2 (n-p+2s)$ where $s$, the signal strength is defined in Eq.~\eqref{eq:s} and given by Eq.~\eqref{eq:signalst}. This dual criterion guarantees that the background model provides a statistically acceptable description of the data without being indistinguishable from noise. III) To enhance the statistical power for testing a subdominant new signal, the null hypothesis $H_0$ must vary slowly in the neighborhood of the best--fit parameters $\hat{\boldsymbol{\eta}}$. In practice, this condition requires that the characteristic curvature scale of the background model be comparable to the typical parameter uncertainties. These uncertainties are estimated numerically during the fitting procedure. Analytically, the parameter covariance matrix is defined as the inverse of the Hessian matrix of the chi--squared function, evaluated at the best--fit point \cite{fisher1925theory}:
\begin{equation}\label{eq:covariance}
C_{\alpha\beta}
=
\left[
\left.
\frac{1}{2}
\frac{\partial^2 \chi^2}{\partial \eta_\alpha\,\partial \eta_\beta}
\right|_{\hat{\boldsymbol{\eta}}}
\right]^{-1}_{\alpha\beta} \, .
\end{equation}
It is convenient to reparametrize the covariance matrix in terms of the marginal parameter uncertainties and their correlation coefficients,
\begin{equation}
C_{\alpha\beta}
=
\sigma_{\eta_\alpha}\,
\sigma_{\eta_\beta}\,
\rho_{\alpha\beta},
\qquad
|\rho_{\alpha\beta}| < 1,
\label{eq:covariance_reparametrization}
\end{equation}
where $\sigma_{\eta_\alpha}^2 = C_{\alpha\alpha}$ and $\rho_{\alpha\beta}$ encodes the correlations between parameters. Since it is complicated to estimate directly the scale of variation of the background model with respect to the covariance matrix in high dimensional parameter space, in order to enforce the third condition, a penalty term can be added to any new fitting function. It is given by
\begin{equation}\label{eq:regular}
\chi^2_{\rm reg} = \lambda (\eta_\alpha - \hat{\eta}_\alpha) \, C^{-1}_{\alpha\beta} \, (\eta_\beta - \hat{\eta}_\beta) \, ,
\end{equation}
where this term, with a suitable scale $\lambda$, is added to the likelihood to regularize the new signal fit to the region of the null-hypothesis uncertainties in parameter space.

When all requirements are met, the null hypothesis $H_0$ is classified as SB: condition (I) provides an acceptable yet not overly good fit to the data, from condition (II) it is clearly distinguishable from random noise, and (III) is locally stable in parameter space so that the background can be reliably linearized with respect to its free parameters near $\hat{\eta}$.

\subsection{HM Hypothesis}
We consider the alternative hypothesis $H_1$, under which the data are described by
\[
\xi_i(\boldsymbol{\theta},\boldsymbol{\eta})
=
b_i(\boldsymbol{\eta}) + \epsilon_i(\boldsymbol{\theta}),
\]
where $\boldsymbol{\theta} = \{\theta_i\}_{i=1}^{p_1}$ denotes the parameters of a putative new (subdominant) signal and
$\boldsymbol{\eta}$ are the background parameters restricted to the allowed region defined in the previous section.
We assume that $H_0$ is nested within $H_1$, such that $\epsilon_i(\boldsymbol{\theta})\equiv0$ recovers $H_0$ for $\boldsymbol{\theta} = 0 $.

To test $H_1$ against $H_0$, we minimize the $\chi^2$ under $H_1$.
Using the linearized expansion of the background model about the background-only best fit
$\hat{\boldsymbol{\eta}}$, the $\chi^2$ under $H_1$ can be written as ~\cite{Fogli:2002pt}
\begin{equation}
\label{eq:chi2_H1}
\chi^2(\boldsymbol{\eta},\boldsymbol{\theta})
=
\sum_{i=1}^n
\frac{
\big(
\delta x_i
-
\sum_\alpha J_{i\alpha}(\eta_\alpha-\hat\eta_\alpha)
-
\epsilon_i(\boldsymbol{\theta})
\big)^2
}{\sigma_i^2},
\end{equation}
where the background-only residuals are
\begin{equation}
\delta x_i
=
x_i - b_i(\hat{\boldsymbol{\eta}}),
\end{equation}
and
$
J_{i\alpha}
$,
the Jacobian matrix is given by
\begin{equation}
J_{i\alpha}
\equiv
\left.
\frac{\partial b_i}{\partial \eta_\alpha}
\right|_{\hat{\boldsymbol{\eta}}}.
\end{equation}
Minimizing Eq.~\eqref{eq:chi2_H1} with respect to $\eta_\beta$ yields
\begin{equation}
\sum_{i=1}^n
\frac{
\big(
\delta x_i
-
\sum_\alpha J_{i\alpha}(\eta_\alpha-\hat\eta_\alpha)
-
\epsilon_i(\boldsymbol{\theta})
\big)
J_{i\beta}
}{\sigma_i^2}
=0.
\end{equation}
Using the background-only best-fit condition
\[
\sum_{i=1}^n \frac{J_{i\beta}\,\delta x_i}{\sigma_i^2}=0,
\]
this equation can be solved for the $\boldsymbol{\theta}$-dependent shift of the parameters,
\begin{equation}
\label{eq:rho_def}
\hat{\hat{\eta}}_{\alpha}(\boldsymbol{\theta})-\hat\eta_\alpha
= -
\sum_{\beta = 1}^{p_0}
C_{\alpha\beta}
\sum_{i=1}^n
\frac{J_{i\beta}\,\epsilon_i(\boldsymbol{\theta})}{\sigma_i^2},
\end{equation}
where
$
C_{\alpha\beta}
$
is the covariance matrix of the background parameters, defined in Eq.~\eqref{eq:covariance}. Equation~\eqref{eq:rho_def} describes the background-induced shift of the  parameters caused by the presence of the signal $\epsilon$.
Substituting this result back into Eq.~\eqref{eq:chi2_H1}, the minimized chi-square under $H_1$ becomes
\begin{equation}
\label{eq:chimin1}
\hat{\hat{\chi}}^2(\boldsymbol{\theta})
=
\sum_{i=1}^n
\big( \frac{\delta x_i - \delta\epsilon_i(\boldsymbol{\theta}) }{\sigma_i} \big)^2,
\end{equation}
where the effective signal after profiling over the background parameters is
\begin{equation}
\delta\epsilon_i(\boldsymbol{\theta}) \equiv \epsilon_i(\boldsymbol{\theta}) - 
\sum_{j=1}^n  \dfrac{\epsilon_j(\boldsymbol{\theta})}{\sigma_j} \Big[ \sum_{\alpha=1}^{p_0} \sum_{\beta=1}^{p_0}  \dfrac{J_{i\alpha} C_{\alpha\beta} J_{j\beta} }{\sigma_j}\Big] ,
\end{equation}
The quantity $\delta\epsilon_i$ therefore represents the component of the signal that cannot be absorbed by a redefinition of the background parameters. Finally from Eq~.(\ref{eq:regular}) we add regularization term to the chi square to insure that the background lineraztion assumption is satisfied:
\begin{equation}
\chi^2(\boldsymbol{\theta}) \equiv \hat{\hat{\chi}}^2(\boldsymbol{\theta}) + \chi^2_{\rm reg}
\end{equation} 
The best-fit value $\hat{\boldsymbol{\theta}}$ is determined by solving 
\[
\frac{\partial \chi^2(\boldsymbol{\theta})}{\partial \theta_\alpha} = 0.
\]
for all the parameters of $\theta_\alpha$.

In order to claim that an alternative hypothesis is HM, two conditions must be satisfied. First, we control the quality of the fit by imposing the condition $Z>-2$, which suppresses the risk of overfitting. Second, when the minimized statistic $\chi^2(\hat{\boldsymbol{\theta}})$ under the alternative hypothesis is smaller than the corresponding $\chi^2(\hat{\boldsymbol{\eta}})$ from the background-only model, we must assess the statistical significance of this improvement. 
To distinguish between random fluctuations and genuine signals in the data, we employ the likelihood ratio test defined by:
\begin{equation}
\mathrm{t} \equiv \chi^2_{\hat{\boldsymbol{\eta}}} - \chi^2_{\hat{\boldsymbol{\theta}}} \, ,
\end{equation}
By combining Eq.~\eqref{eq:chimin0} and Eq.~\eqref{eq:chimin1}, we derive the test statistic. Under the null hypothesis, the first term corresponds to the sum of squared noise contributions, while $\mathrm{t}$ follows a $\chi^2$ distribution with degrees of freedom equal to the number of additional parameters $\theta_\alpha$ in $H_1$ compared to $H_0$. Depending on the number of additional parameters, we can estimate how unlikely the result of the test statistic is, given that the null hypothesis is true.

\section{Summary and conclusions}
In this study, we have focused on experimental results consistent with background predictions, which have been very common in recent years, while acknowledging the existence of highly motivated (HM) theories that require empirical constraints. The presented procedure offers a simple and efficient approach to evaluate the compatibility of alternative hypotheses with data compared to the background model, in cases where the analytical formula for the background is defined or the numerical values are accessible for computation. While the analysis employs a simplified background model, actual experimental scenarios involve complex background sources from multiple contributions, in which computing the numerical derivative becomes challenging. To address this complexity, it may be possible to use automatic differentiation programming to improve the efficiency and speed of the derivative calculations~\cite{jax2018github}. In cases where the bins are not Gaussian, or where the distribution of the test statistic does not follow the chi-square distribution due to experimental limitations, one has to use a modified version of Eq.~\eqref{eq:chimin1} and determine the distribution of the test statistic through numerical simulations.


After examining these considerations, it must be emphasized that any detected hints can be particularly valuable for future experiment planning. Moreover, in cases showing null compatibility, the same framework can be applied to place constraints on new physics parameters.






\bibliographystyle{elsarticle-harv} 
\bibliography{May6.bib}

@inproceedings{fisher1925theory,
  title={Theory of statistical estimation},
  author={Fisher, Ronald Aylmer},
  booktitle={Mathematical proceedings of the Cambridge philosophical society},
  volume={22},
  number={5},
  pages={700--725},
  year={1925},
  organization={Cambridge University Press}
}

@software{jax2018github,
  author = {James Bradbury and Roy Frostig and Peter Hawkins and Matthew James Johnson and Yash Katariya and Chris Leary and Dougal Maclaurin and George Necula and Adam Paszke and Jake Vander{P}las and Skye Wanderman-{M}ilne and Qiao Zhang},
  title = {{JAX}: composable transformations of {P}ython+{N}um{P}y programs},
  url = {http://github.com/jax-ml/jax},
  version = {0.3.13},
  year = {2018},
}

@article{ParticleDataGroup:2024cfk,
    author = "Navas, S. and others",
    collaboration = "Particle Data Group",
    title = "{Review of particle physics}",
    doi = "10.1103/PhysRevD.110.030001",
    journal = "Phys. Rev. D",
    volume = "110",
    number = "3",
    pages = "030001",
    year = "2024"
}

@article{Fogli:2002pt,
    author = "Fogli, G. L. and Lisi, E. and Marrone, A. and Montanino, D. and Palazzo, A.",
    title = "{Getting the most from the statistical analysis of solar neutrino oscillations}",
    eprint = "hep-ph/0206162",
    archivePrefix = "arXiv",
    doi = "10.1103/PhysRevD.66.053010",
    journal = "Phys. Rev. D",
    volume = "66",
    pages = "053010",
    year = "2002"
}

@article{Wilks:1938dza,
    author = "Wilks, S. S.",
    title = "{The Large-Sample Distribution of the Likelihood Ratio for Testing Composite Hypotheses}",
    doi = "10.1214/aoms/1177732360",
    journal = "Annals Math. Statist.",
    volume = "9",
    number = "1",
    pages = "60--62",
    year = "1938"
}

@article{vanDyk:2023tqz,
    author = "van Dyk, David and Lyons, Louis",
    title = "{How to Incorporate Systematic Effects into Parameter Determination}",
    eprint = "2306.05271",
    archivePrefix = "arXiv",
    primaryClass = "hep-ex",
    month = "6",
    year = "2023"
}

@article{Lyons:2017zlp,
    author = "Lyons, Louis",
    title = "{A Paradox about Likelihood Ratios?}",
    eprint = "1711.00775",
    archivePrefix = "arXiv",
    primaryClass = "physics.data-an",
    month = "11",
    year = "2017"
}

@article{Ranucci:2005ep,
    author = "Ranucci, Gioacchino",
    title = "{Likelihood scan of the Super-Kamiokande I time series data}",
    eprint = "hep-ph/0511026",
    archivePrefix = "arXiv",
    doi = "10.1103/PhysRevD.73.103003",
    journal = "Phys. Rev. D",
    volume = "73",
    pages = "103003",
    year = "2006"
}

@article{Fowlie:2021gmr,
    author = "Fowlie, Andrew and Hoof, Sebastian and Handley, Will",
    title = "{Nested Sampling for Frequentist Computation: Fast Estimation of Small p-Values}",
    eprint = "2105.13923",
    archivePrefix = "arXiv",
    primaryClass = "physics.data-an",
    doi = "10.1103/PhysRevLett.128.021801",
    journal = "Phys. Rev. Lett.",
    volume = "128",
    number = "2",
    pages = "021801",
    year = "2022"
}

@article{Blennow:2013oma,
    author = "Blennow, Mattias and Coloma, Pilar and Huber, Patrick and Schwetz, Thomas",
    title = "{Quantifying the sensitivity of oscillation experiments to the neutrino mass ordering}",
    eprint = "1311.1822",
    archivePrefix = "arXiv",
    primaryClass = "hep-ph",
    doi = "10.1007/JHEP03(2014)028",
    journal = "JHEP",
    volume = "03",
    pages = "028",
    year = "2014"
}

@article{Davies:1987zz,
    author = "Davies, Robert B.",
    title = "{Hypothesis testing when a nuisance parameter is present only under the alternative}",
    doi = "10.1093/biomet/74.1.33",
    journal = "Biometrika",
    volume = "74",
    pages = "33--43",
    year = "1987"
}

@article{Gross:2010qma,
    author = "Gross, Eilam and Vitells, Ofer",
    title = "{Trial factors for the look elsewhere effect in high energy physics}",
    eprint = "1005.1891",
    archivePrefix = "arXiv",
    primaryClass = "physics.data-an",
    doi = "10.1140/epjc/s10052-010-1470-8",
    journal = "Eur. Phys. J. C",
    volume = "70",
    pages = "525--530",
    year = "2010"
}

@article{Cowan:2010js,
    author = "Cowan, Glen and Cranmer, Kyle and Gross, Eilam and Vitells, Ofer",
    title = "{Asymptotic formulae for likelihood-based tests of new physics}",
    eprint = "1007.1727",
    archivePrefix = "arXiv",
    primaryClass = "physics.data-an",
    doi = "10.1140/epjc/s10052-011-1554-0",
    journal = "Eur. Phys. J. C",
    volume = "71",
    pages = "1554",
    year = "2011",
    note = "[Erratum: Eur.Phys.J.C 73, 2501 (2013)]"
}

@article{Cousins:2018tiz,
    author = "Cousins, Robert D.",
    title = "{Lectures on Statistics in Theory: Prelude to Statistics in Practice}",
    eprint = "1807.05996",
    archivePrefix = "arXiv",
    primaryClass = "physics.data-an",
    month = "7",
    year = "2018"
}

@article{Junk:2020azi,
    author = "Junk, Thomas R. and Lyons, Louis",
    title = "{Reproducibility and Replication of Experimental Particle Physics Results}",
    eprint = "2009.06864",
    archivePrefix = "arXiv",
    primaryClass = "physics.data-an",
    reportNumber = "FERMILAB-PUB-20-649-ND",
    doi = "10.1162/99608f92.250f995b",
    month = "9",
    year = "2020"
}






\end{document}